\documentclass[pra,superscriptaddress,showpacs,twocolumn,amsmath]{revtex4}
\usepackage{graphicx}

\newcommand{\beq}{\begin{equation}}
\newcommand{\eeq}{\end{equation}}
\newcommand{\bea}{\begin{eqnarray}}
\newcommand{\eea}{\end{eqnarray}}
\newcommand{\bml}{\begin{subequations}}
\newcommand{\eml}{\end{subequations}}
\newcommand{\ba}{\begin{array}}
\newcommand{\ea}{\end{array}}
\newcommand{\bc}{\begin{center}}
\newcommand{\ec}{\end{center}}
\newcommand{\commentout}[1]{{}}
\newcommand{\bk}{{\bf k}}

\newcommand{\adag}{a^\dagger}

\newcommand{\bdag}{b^\dagger}

\newcommand{\half}{\hbox{$\frac{1}{2}$}}
\newcommand{\fourth}{\hbox{$\frac{1}{4}$}}

\newcommand{\eq}[1]{(\ref{#1})}

\newcommand{\vol}[1]{{\bf #1}}
\newcommand{\comment}[1]{{}}

\begin{document}

%\large

\title{All Optical Scheme for Strongly Enhanced Production of Dipolar Molecules in the Electro-Vibrational Ground State}
\author{Matt Mackie}
\affiliation{Department of Physics, Temple University, Philadelphia, PA 19122}
\author{Catherine Debrosse}
\affiliation{Department of Physics, Temple University, Philadelphia, PA 19122}
\affiliation {Department of Biology, Temple University, Philadelphia, PA 19122}
\date{\today}
\begin{abstract}
We consider two-color heteronuclear photoassociation of atoms into dipolar molecules in the $J=1$ electro-vibrational ground state, where a free-ground laser couples atoms directly to the ground state and a free-bound laser couples the atoms to an electronically-excited state. This problem raises an interest because heteronuclear photoassociation from atoms to near-ground state molecules is limited by the small size of the target state.  Nevertheless, the addition of the excited state creates a second pathway for creating ground state molecules, leading to quantum interference between direct photoassociation and photoassociation via the excited molecular state, as well as a dispersive-like shift of the free-ground resonance position. Using LiNa as an example, these results are shown to depend on the detuning and intensity of the free-bound laser, as well as the semi-classical size of both molecular states. Despite strong enhancement, coherent conversion to the LiNa electro-vibrational ground state is possible only in a limited regime near the free-bound resonance.
\end{abstract}


\maketitle

\section{Introduction}

As a gas of $N$ bosons is cooled to a temperature $T$ in the nanocold regime, the thermal DeBroglie wavelength of the atoms, $\Lambda_D\propto1/\sqrt{T}$, gets longer and longer until, eventually, it is of the order of the interparticle spacing, $\rho\Lambda_D^3\sim1$ (where $\rho$ is the particle density), and the atoms condense into a single quantum state. For alkali-metal atoms, Bose-Einstein condensation occurs around 100~nK, and is achieved by laser and evaporative cooling atoms that begin at oven temperatures $\sim600$~K~\cite{COR02}. Unfortunately, laser cooling depends on the existence of a closed few-level system, which does not exist for molecules, making quantum degeneracy harder to achieve. Nevertheless, due to large phase space density~\cite{BUR96}, photoassociation of Bose-condensed atoms can be highly efficient, providing an end run around the need for laser cooling of molecules. In the heteronuclear case~\cite{SHA99,MAR99}, photoassociation can thus serve as a bulk source of dipolar molecules which, in turn, can be used in quantum computing applications~\cite{DEM02} and simulations of condensed-matter systems~\cite{BUE07}.

In photoassociation~\cite{LET93,THO87,WEI99}, a colliding pair of atoms absorbs a photon and makes a transition from the two-atom continuum to an bound molecular state. Since one-photon transitions to the homonuclear ground state are dipole forbidden, these free-bound transitions occur to an electronically-excited state that undergoes spontaneous decay, typically on a time scale of tens of nanoseconds. To allow for fundamental studies and practical applications, a second laser is needed to drive population to the ground state~\cite{JON97,WYN00,VAR97}. However, Raman photoassociation is limited in practice by unwanted dissociation~\cite{KOS00}, as well as the large frequency difference between the lasers that is required to target vibrational stables. So far, free-bound-ground photoassociation experiments have produced only a relatively small number of molecules in metastable high-lying levels~\cite{WYN00}.

\begin{figure}[t]
\centering
\includegraphics[width=8.5cm]{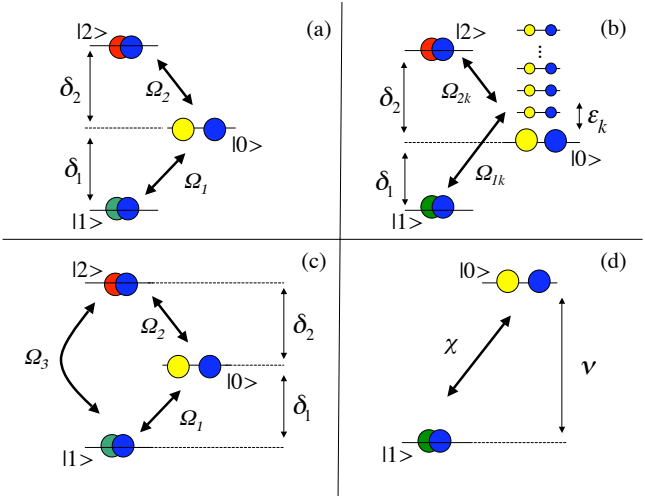}
\caption{(a) Ground-free-bound photoassociation of a heteronuclear conensate, including (b) dissociation to noncondensate modes. (c) Treating the noncondensate modes in the steady-state approximation leads to an effective coupling between the two molecular states. (d) Similarly, for large free-bound detuning, an effective two-level system emerges.}
\label{FEWL}
\end{figure}

We therefore consider photoassociation of a heteronuclear condensate, where the permanent electric dipole moment of the daughter molecule enables transitions directly to the ground electro-vibrational state. High phase space density not withstanding, excessive laser power is required in a one-photon scheme due to the the small size of the molecular state relative to the atom-pair state. A second photoassociation laser is then added to drive free-bound transitions to an electronically-excited molecular state, creating the ``ground-free-bound" system shown in Fig.~\ref{FEWL}. The addition of the second photoassociation laser creates an alternative path for reaching the electro-vibrational ground state, leading to quantum interference and a dispersive-like shift of the free-ground resonance position that depend on the free-bound laser intensity and detuning, as well as the size of both molecular states. Nevertheless, strong enhancement does not guarantee coherent conversion to the electro-vibrational ground state, which occurs only for limited detunings near free-bound resonance. This scheme is analogous to laser-induced autoionization~\cite{FAN61}, and Feshbach-enhanced photoassociation~\cite{COU98,MAC08,PEL08,DEB09}.

This Article is outlined as follows. In Sec.~\ref{MODEL}, the model for ground-free-bound photoassociation of a Bose-Einstein condensate of heteronuclear atoms is introduced, including photodissociation to noncondensate atom pairs. Next, a steady-state solution is developed for the noncondensate pair amplitude, leading to a cross-molecular coupling between the two, otherwise uncoupled, molecule states. Also, the rate equation is derived for laser-enhanced photoassociation using the equivalent of the density matrix from quantum optics, and quantum interference is discussed qualitatively. In Sec.~\ref{RESULTS}, the detuning and intensity position of the inteference node and lightshift are detailed, as well as enhanced saturation intensties, using LiNa as an example. Lastly, coherent conversion to the electro-vibrational ground state is shown to take place in a finite window near the free-bound resonance. We conclude with Sec.~\ref{CONCLUSION}.

\section{Steady-State Model}
\label{MODEL}

Consider $N_1$ atoms of mass $m_1$ and $N_2$ atoms of mass $m_1$ that have each Bose-condensed into the zero-momentum state $|0\rangle$, with $N=N_1+N_2$ as the total number of atoms. The free-ground photoassociation laser then destroys an atom from each species and creates a dipolar molecule of mass $m_3=m_1+m_2$ in the  $J=1$ rotational state of the electro-vibrational ground state, $|1\rangle$, where transtions to the absolute ground state are dipole forbidden. The natural linewidth for this state is $\Gamma_{01}$, the atom-molecule Rabi coupling is $\Omega_1$, and the detuning $\delta_{01}<0$ indicates an open dissociation channel. Similarly, the free-bound photoassociation laser creates a molecule in the excited state $|2\rangle$ with natural linewidth $\Gamma_{02}$, where the atom-molecule coupling is $\Omega_2$ and the detuning $\delta_{02}>0$ denotes an open dissociation channel.  As per Fig.~\ref{FEWL}(a), this is the cascade system familiar from few-level quantum optics with the intermediate state initially occupied. In the language of second quantization,  annihilation of an atom (molecule) from the $i$-th atomic (molecular) condensate is represented by the operator $a_i$ ($b_i$). The Hamiltonian for the condensate modes is
\bml
\bea
\frac{H_0}{\hbar}&=&
  \sum_i\left[(-1)^i\delta_{0i}-i\Gamma_{0i}/2)\bdag_i b_i\right]
\nonumber\\&&
    -\frac{1}{2\sqrt{N}}\sum_i\Omega_i\left[a_1 a_2\bdag_i +b_i\adag_2\adag_1\right].
\label{H_BEC}
\eea

To be more precise, molecular dissociation to noncondensate levels should be included, as per Fig.~\ref{FEWL}(b). This situation arises because a condensate molecule need not dissociate back to the atomic condensate, but may just as well create a pair of atoms with equal-and-opposite momentum, since total momentum is conserved. So-called rogue \cite{KOS00,JAV02}, or unwanted \cite{GOR01}, dissociation to noncondensate modes therefore introduces the operators $a_{i,\pm\bk}$, along with the kinetic energy of a dissociated pair $\hbar\varepsilon_\bk=\hbar^2k^2/2\mu$ of reduced mass $1/\mu=1/m_1+1/m_2$. The noncondensate Hamiltonian is
\bea
\frac{H_1}{\hbar}&=&
  \sum_{\bk}\varepsilon_\bk\left(\adag_{1,\bk}a_{1\bk}+\adag_{2,\bk}a_{2\bk}\right)
\nonumber\\&&
    -\frac{1}{2\sqrt{N}}\sum_i\sum_\bk \Omega_i f_{i,\bk}(a_{1,\bk}a_{2,-\bk}\bdag_i 
        +b_i\adag_{2,-\bk}\adag_{1,\bk}),
\nonumber\\
\label{H_ROGUE}
\eea
\label{H_FULL}
\eml
where $f_{i,\bk}$ accounts for the wavevector dependence of the $i$-th coupling to the noncondensate modes. Note that the free term in $H_1$ is an approximation that yields the correct mean-field equations to lowest order in the BBGKY hierarchy of correlation functions.

To obtain these mean-field equations, the Heisenberg equation of motion for a given operator, $i\hbar\dot{x}=[x,H]$, are derived from the Hamiltonian $H=H_0+H_1$ given by Eqs.~\eq{H_FULL} and the operators are declared $c$-numbers. In a minimalist model, $x$ represents either the atomic amplitude $a_i$  (molecular amplitude $b_i$), or the anomalous density, $A_\bk=a_{1,\bk} a_{2,-\bk}$, which arises from photodissociation to noncondensate modes. The summation over $\bk$ implicit to the Hamiltonian is then converted to an integral over frequency, introducing the frequency $\omega_\rho=\hbar\rho^{2/3}/2\mu$. The resulting mean-field equations are
\bml
\bea
i\dot{a}_1 &=& -\half a_2^*(\Omega_1 b_1+\Omega_2 b_2), 
\\
i\dot{a}_2 &=& -\half a_1^* (\Omega_1 b_1+\Omega_2 b_2), 
\\
i\dot{b}_i &=&  \left[(-1)^i\delta_{0i}-i\Gamma_{0i}/2\right] b_i-\half\Omega_i a_1a_2
\nonumber\\&&
  -\half\xi_i\!\int\!d\varepsilon\sqrt{\varepsilon}\,f_i(\varepsilon)A(\varepsilon),
\label{BOSE_BDOT}
\\
i\dot{A}(\varepsilon) &=&\varepsilon A(\varepsilon)
    -\half\,{\textstyle\sum_i}\Omega_i\,f_i(\varepsilon)b_i,
\eea
\label{BOSE_EQM}
\eml
where the amplitudes have been scaled by a factor of $\sqrt{N}$ and are now of order unity, the $i$th atom-molecule coupling is Bose-enhanced, $\Omega_i\propto\sqrt{\rho}$, and the rogue coupling is $4\pi\xi_i=\Omega_i/\omega_\rho^{3/2}$. The frequency dependence of the $i$th atom-molecule coupling is modeled with a Gaussian, $f_i(\varepsilon)=\exp(-\varepsilon^2/2\beta_i^2)$, where the width of the coupling, $\beta_i$, is determined by the semi-classical size of the $i$-th molecule state, $\beta_i=\hbar/(2\mu L_i^2)$.

Borrowing from quantum optics, the simplest nontrival way to include noncondensate modes is to treat the anomalous amplitude in the steady-state approximation, $\dot{A}\approx0$. The system then evolves according to
\bml
\bea
i\dot{a}_1 &=& -\half a_2^*(\Omega_1 b_1+\Omega_2 b_2), 
\\
i\dot{a}_2 &=& -\half a_1^* (\Omega_1 b_1+\Omega_2 b_2), 
\\
i\dot{b}_1 &=& -(\delta_1+i\Gamma_1/2) b_1
  -\half\Omega_1 a_1a_2-\half\Omega_3b_2,
\\
i\dot{b}_2 &=& (\delta_2-i\Gamma_2) b_2
  -\half\Omega_2 a^2-\half\Omega_3b_1,
\eea
\label{VCAS_EQM}
\eml
where $\delta_i=\delta_{0i}-(-1)^i\sigma_{0i}$ and $\Gamma_i=\Gamma_{0i}+\gamma_i$. The shift of laser resonance is $\sigma_{0i}=\Re[\Sigma_i]$, and the photodissociation rate is $\gamma_i=\Im[\Sigma_i]$, where
\bml
\beq
\Sigma_i=\lim_{\omega\rightarrow0}\fourth\Omega_i\xi_i
  \int d\varepsilon\sqrt{\varepsilon}\frac{f_i^2(\varepsilon)}{(\varepsilon-\omega)}.
\eeq
Also, the shared dissociation continuum leads to an effective coupling between the otherwise uncoupled molecular states,
\beq
\Omega_3=
 \frac{\Omega_1\Omega_2}{4\pi\omega_\rho^{3/2}}\,
   \Re\left[\lim_{\omega\rightarrow0}\int d\varepsilon
     \sqrt{\varepsilon}\frac{f_1(\varepsilon)f_2(\varepsilon)}{(\varepsilon-\omega)}\right].
\eeq
\eml
Physically, the shared continuum acts like a virtual state that enables transitions between the two, otherwise uncoupled, photoassociation targets. 

Next, assume that the free-bound laser is far-off resonance, whereby the excited amplitude $b_2$ is treated in steady-state as well. The system is now described by an effective two-level (three-mode) model [Fig.~\ref{FEWL}(d)]:
\bml
\bea
i\dot{a_1}&=&-\frac{\Omega_2^2}{4\delta_2} |a_2|^2a_1
  -\half\chi a_2^*b_1,
\\
i\dot{a_2}&=&-\frac{\Omega_2^2}{4\delta_2} |a_1|^2a_2
  -\half\chi a_1^*b_1,
\\
i\dot{b}_1&=&-(\nu+i\Gamma/2) b_1-\half\chi a_1a_2,
\eea
\label{TWOL}
\eml
where the two-photon atom-molecule coupling is
$\chi=\Omega_1+\half{\cal L}\delta_2\Omega_2/\Omega_3$, the Stark-shifted detuning is $\nu=\delta_1+\fourth{\cal L}\delta_2$, and the effective damping is $\Gamma=\Gamma_1+\fourth{\cal L}\Gamma_2$, where ${\cal L}=\Omega_3^2/[\delta_2^2+\Gamma_2^2/4]$. Since $\Gamma_2\gg[\Omega_2^2/(4\delta_2)] |a_i|^2$, we neglect the resonant mean-field shift compared to the natural linewidth of the excited molecular state, except to define the resonant intercondensate scattering length, $\rho^{1/3}a_r=-\Omega_2^2/(8\pi\omega_\rho\delta_2)$.

To derive an expression for the rate constant, we define the atomic and molecular probabilities, $P_i=a_i^*a_i$  and $P_0=2b_1^*b_1$, as well as the atom-molecule ``coherence" $C_{0i}=a_ib_1^*$. Making extensive use of the product rule, e.g., $i\dot C_{0i}=ia_ia_i\dot b_1^*+2ia_i\dot{a}_ib_1^*$, we have
\bml
\bea
i\dot P_i &=& \chi(C_{0i}-C_{0i}^*), \\
i\dot P_0 &=& -i\Gamma P_m -\half\chi\sum_i(C_{0i}-C_{0i}^*), \\
i\dot C_{0i} &=& (\nu-i\Gamma/2)C_{0i}
  +\half \chi P_i(P_i-2P_0).
\eea
\label{GEN_RATE}
\eml
Equations~\eq{GEN_RATE} are equivalent to the density matrix from quantum optics.
Solving Eqs.~\eq{GEN_RATE} in the reservoir approximation, $P_i\sim1$ and $P_0\ll1$, and for a steady-state coherence, $\:\dot{C}_{0i}\approx0$, the rate equation for the probability of the $i$th atomic condensate is given by
\bml
\beq
\dot{P}_i=\frac{1}{4}\,\frac{\chi^2/\Gamma}{\nu^2+\Gamma^2/4}\,P_1P_2.
\label{P_DOT}
\eeq
The rate equation~\eq{P_DOT} defines the rate constant for heteronuclear photoassociation
\beq
\rho K=\frac{1}{4}\,\frac{\chi^2/\Gamma}{\nu^2+\Gamma^2/4}\,.
\label{RATE_CON}
\eeq
\eml

Evidentally, as a function of the laser detuning from the free-bound resonance, $\delta_2$, the laser resonant rate constant $K$ is suppressed below the rate for photoassociation alone for $\chi<\Omega_1$, essentially vanishing for $\chi\approx0$, and is enhanced above the rate for photoassociation alone for $\chi>\Omega_1$. Borrowing intuition from quantum optics, once the two molecular states are coupled by the shared photodissociation continuum, this suppression and enhancement is due to quantum interference between photoassociation directly to the electro-vibrational ground state and photoassociation to the electro-vibrational ground state via the electro-excited molecular state. 

\section{Explicit Results}
\label{RESULTS}
\commentout{
Generally, this quantum interference is tunable according to both the detuning and intensity of the free-bound laser.

Borrowing from Ref.~\cite{PEL08}, the unit-intensity rate constant for the $J=1$ state in the electro-vibrational ground state of LiNa is $K_1=5\times10^{-19}$~cm$^3$/s, and the natural linewidth is $\Gamma_{01}=12\times2\pi$~mHz, so that $\Omega_1=0.98\times2\pi$~mHz for $I_1=1$~kW/cm$^2$ and $\rho=10^{12}$~cm$^{-3}$.
}

For concreteness, we consider explicit numbers for LiNa. The rate constant for resonant free-ground photoassociation alone is
\beq
\rho K_1=\frac{\Omega_1^2}{\Gamma_{01}+\gamma_1}.
\label{BARE_RATE_CON}
\eeq
In the low intensity limit $\gamma_1\ll\Gamma_{01}$, so that $\Omega_1=\sqrt{\rho K_1\Gamma_0}$. Borrowing from Ref.~\cite{PEL08}, the unit-intensity rate constant for 100~nK~\cite{TEMP_NOTE} transitions to the $J=1$ rotational state in the electro-vibrational ground state of LiNa is $K_1=5.5\times10^{-15}$~cm$^3$/s. Estimating the natural linewidth $\Gamma_{01}=12\times2\pi$~mHz, the free-ground Rabi coupling is $\Omega_1=3.2\times2\pi$~mHz for $I_1=1$~W/cm$^2$ and $\rho=10^{12}$~cm$^{-3}$. The free-bound coupling for the electro-excited molecular state, $\Omega_2=\Omega_{02}\sqrt{I/I_{02}}$, is estimated by mass scaling those for homonuclear Li~\cite{MAC08,PRO03}: $\Omega_{02}=44\times2\pi$~kHz/(W/cm$^2$)$^{1/2}$, $I_{02}=43$~W/cm$^2$, $L_2=133a_0$, and $\Gamma_{02}=12\times2\pi$~MHz. Finally, for Gaussian $f_i(\varepsilon)$, the cross-molecular coupling is $\Omega_3=\eta\Omega_1\Omega_2\sqrt{\beta/\omega_\rho}/(4\pi\omega_\rho)$, where $\eta$ is a dimensionless constant of order unity leftover from integrating the two Gaussians, and $\beta=\hbar/(2\mu)L^2$ with $L=\sqrt[4]{L_1^4+L_2^4}$. Hence, away from the free-bound resonance and on lightshifted resonance ($\nu=0$), the rate constant is then
\beq
\rho K=
  \frac{\Omega_1^2\left[1+\eta\left({\displaystyle \frac{1}{\rho^{1/3}L}}\right)
    \left({\displaystyle\frac{\Omega_2^2}{8\pi\omega_\rho\delta_2}}\right)\right]^2}
  {\Gamma_1+\left[\eta
    \left({\displaystyle\frac{1}{\rho^{1/3}L}}\right)
      \left({\displaystyle\frac{\Omega_2^2}{8\pi\omega_\rho\delta_2}}\right)
        \left({\displaystyle\frac{\Omega_1}{\Omega_2}}\right)\right]^2
          \Gamma_2}\,,
\label{RES_RATE_CON}
\eeq
where $\Gamma_1\approx\Gamma_{01}$ for free-ground intensities herein.

\subsection{Destructive Interference}
Considering the node, for fixed free-bound laser intensity photoassociation ceases for the free-bound detuning
\bml
\beq
\frac{\delta_{N2}}{\Gamma_s}=-\frac{\eta}{8\pi}\,
  \frac{\Omega_2^2}{\omega_\rho\Gamma_s}\,\rho^{1/3}L.
\label{DET_NODE}
\eeq 
On the other hand, for fixed laser detuning and $\Omega_i=\Omega_{0i}\sqrt{I_i/I_{0i}}$, where $I_{0i}$ is a characteristic intensity for the given transition, then the intensity position of the destructive interference node is 
\beq
\frac{I_{N2}}{I_{02}}=
  -\frac{8\pi}{\eta}\,\frac{\omega_\rho\delta_2}{\Omega_{02}^2}\,\rho^{1/3}L\,,
  \label{I_NODE}
\eeq
\eml
This node exists in addition to any nodes in the solution to the radial Schr\"odinger equation~\cite{KOS00}, which occur for arbitrary intensity. 

\begin{figure}[b]
\centering
\includegraphics[width=8.5cm]{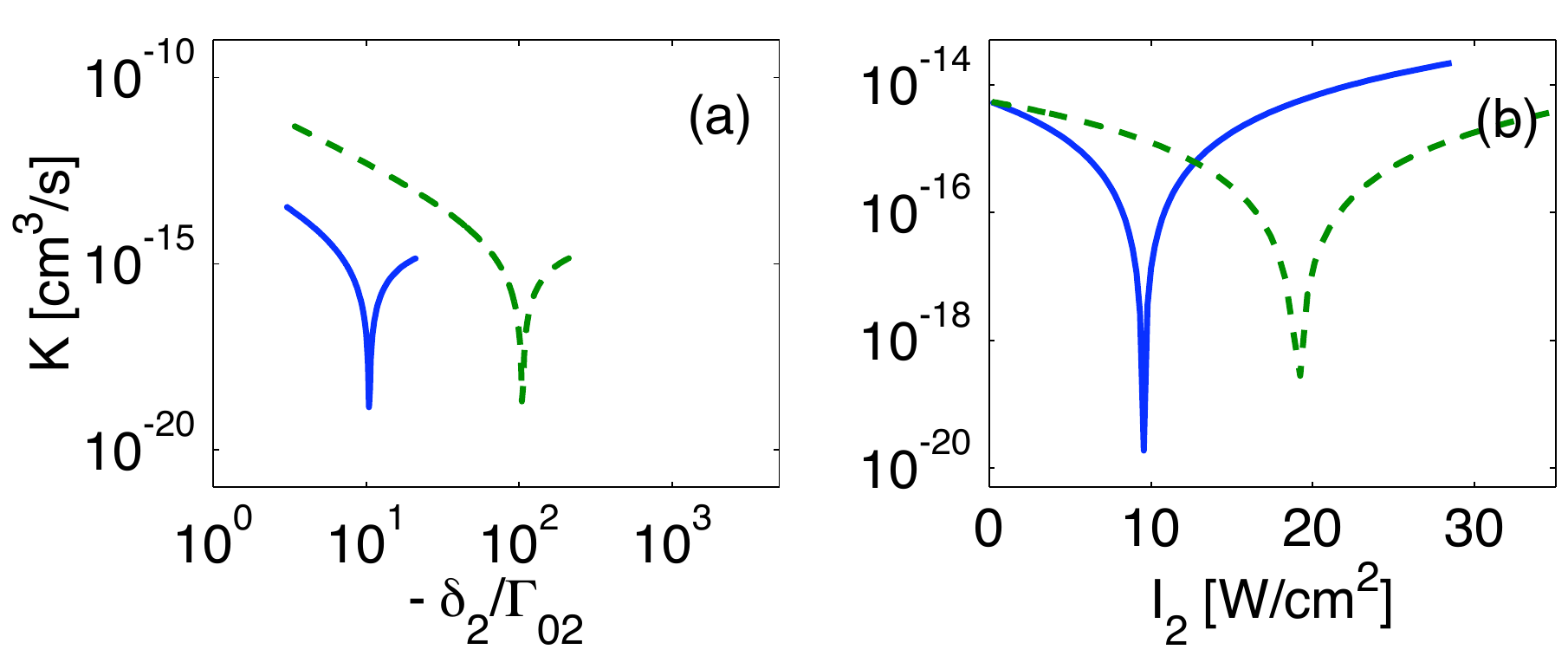}
\caption{Laser-enhanced photoassociation of a joint Li-Na condensate. Here the free-ground laser is on lightshifted resonance $\nu=0$, and $I_1=1$~W/cm$^2$. (a) For fixed free-bound intensities $I_2/I_1=10$ (solid line) and 100 (dashed line), quantum interference is tunable according to the free-bound detuning. (b) On the other hand, for fixed detuning $-\delta_2/\Gamma_{02}=1$ (solid line) and 2 (dashed line), quantum interference is tunable according to the free-ground intensity.}
\label{NODE}
\end{figure}

Results for LiNa are shown in Fig.~\ref{NODE} for modest free-ground intensity $I_1=1$~W/cm$^2$. For fixed free-bound intensity $I_2/I_1=10$ (100), photoassociation ceases at the detuning $\delta_{N2}/\Gamma_{02}\sim10$~(100), as per Eq.~\eq{DET_NODE} and Fig.~\ref{NODE}(a). On the other hand, for fixed free-ground detuning $-\delta_2/\Gamma_{02}=10$~(20), cessation occurs at the intensity $I_2/I_{02}=0.23$~(0.46), as per Eq.~\eq{DET_NODE} and Fig.~\ref{NODE}(b). Although independent of free-ground intensity and detuning, the position of a given node does depend on the size of the free-ground molecules state, as well as the size of the free-bound molecular state.

\subsection{Constructive Interference}

Enhanced or not, the rate constant is linear for low free-ground intensity and saturates to a constant value for large free-ground intensity. Absent the free-bound coupling [Eq.~\eq{BARE_RATE_CON}], saturation occurs at an intensity $I_{01}=\Gamma_{01}/(\gamma_{01}/I_{01})$, which is inconveniently large at best since the free-ground coupling is exceptionally weak, e.g., for LiNa at 100~nK the saturation intensity is $I_{01}\approx2.2$~MW/cm$^2$. However, the laser-enhanced free-ground rate constant is nearly $10^{-11}$~cm$^3$/s [Fig~\ref{NODE}(a)], representing an enhancement of more than three orders of magnitude over one-color alone. Quantum enhancement via the shared continuum might therefore bring saturation of free-ground transitions closer to the realm of observability. Saturation sets in when the second term in the denominator of Eq.~\eq{RES_RATE_CON} is of the order of $\Gamma_1$. The saturated rate constant is then
\bml
\beq
\rho K=
  \left[1+\frac{1}{\eta}\,\rho^{1/3}L\,
    \left(\frac{8\pi\omega_\rho\delta_2}{\Omega_2^2}\right)\right]^2
  \frac{\Omega_2^2}{\Gamma_2}\,,
\eeq 
and the enhanced free-ground saturation intensity is
\beq
\frac{I_{01}'}{I_{01}}=\left[\frac{1}{\eta}\,(\rho^{1/3}L)
        \left(\frac{8\pi\omega_\rho\delta_2}{\Omega_2^2}\right)
      \left(\frac{\Omega_2}{\Omega_{01}}\right)\right]^2
   \frac{\Gamma_1}{\Gamma_2},
\eeq
\eml
Near the free-bound resonance ($\delta_2\sim0$), the rate constant saturates at the rate constant for free-bound photoassociation alone, $\rho K=\Omega_2^2/\Gamma_2$.

\begin{figure}[b]
\centering
\includegraphics[width=8.5cm]{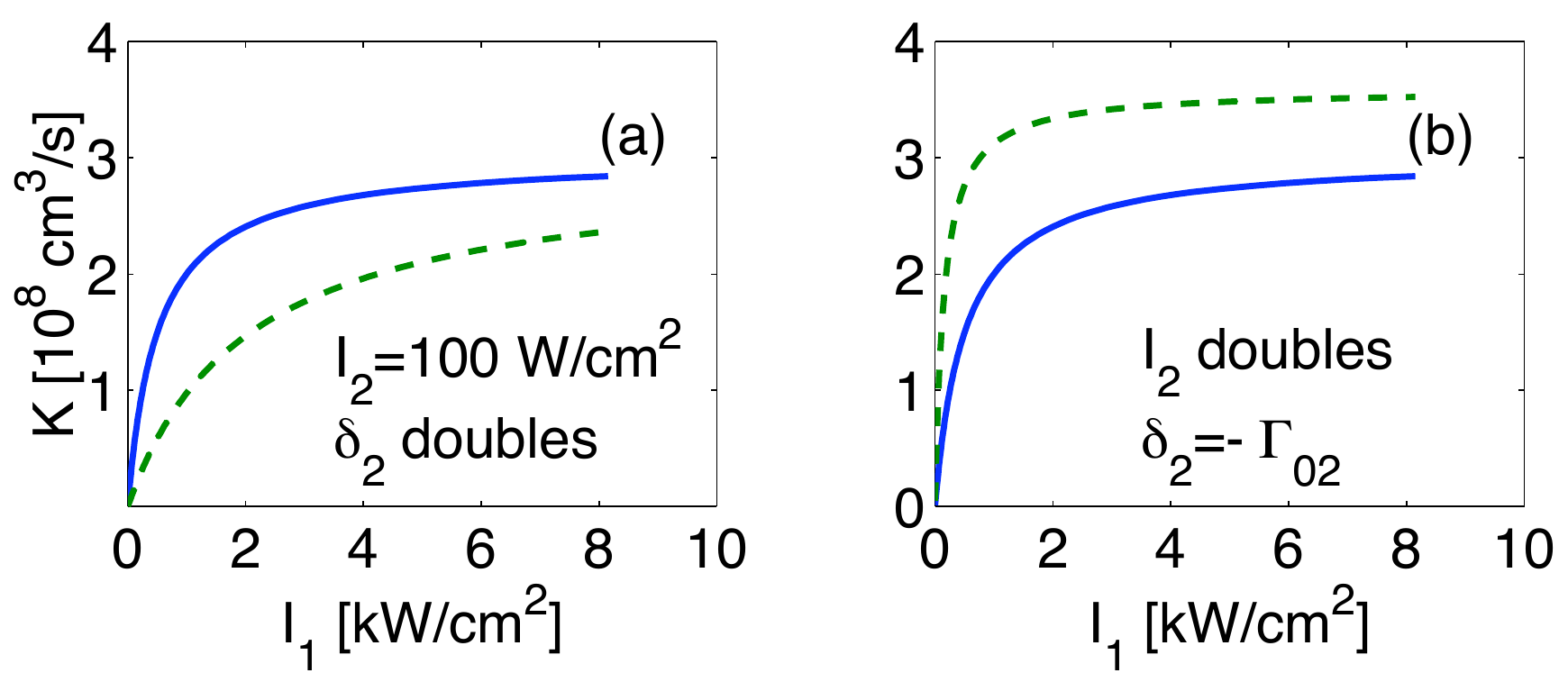}
\caption{Saturation in laser-enhanced free-ground photoassociation of an Li-Na condensate, where the free-ground laser intensity is $I_1=1$~W/cm$^2$. (a) Dependence of saturation on the detuning of the free-bound laser, where $-\delta_2/\Gamma_{02}=1$ (solid line) and 2~(dashed line) for fixed free-bound laser intensity $I_2/I_1=100$. (b) Dependence of saturation on free-bound intensity, where $I_2/I_1=100$ (solid line) and 200 (dashed line) for fixed free-bound detuning $-\delta_2/\Gamma_{02}=1$.}
\label{SATR8}
\end{figure}

For a LiNa system at free-bound detuning on the edge of validity for the present model, $\delta_2/\Gamma_{02}=-1$, and a strong free-bound intensity $100$~W/cm$^2$, the enhanced saturation intensity is $I_{01}\sim2$~kW/cm$^2$. Although still somewhat large, this result is an improvement to $I_{01}$ of three orders of magnitude. Further results are shown in Fig.~\ref{SATR8}, illustrating the dependence on free-bound detuning and intensity. Furthermore, based on intuition from a combination of photoassociation and Feshbach resonances~\cite{MAC08}, the model probably remains valid for much smaller detunings, which brings saturation even further into the observable realm. 

\subsection{Dispersive-like Lightshift}

Whereas the lightshift of laser resonance arising from the free-ground coupling, $\sigma_{01}=\Re[\Sigma_1]$, is strictly to the red~\cite{JON97,GER01,FED96}, the cross-molecular coupling, i.e., the additional bound state, leads to an anomalous lightshift of laser resonance, $\sigma_1=\sigma_{01}+{\cal L}\delta_2$, that can be either blue or red:
\beq
\sigma_1=\sigma_{01}\left[1+\eta^2\,\frac{L_1}{L}\,\left(\frac{1}{\rho^{1/3}L}\right)\,
  \left(\frac{\Omega_2^2}{8\pi\omega_\rho\delta_2}\right)\right].
\eeq
Qualitatively, free-ground photoassociation dominates far from the free-bound resonance, and the lighthsift is therefore to the red ($\sigma_1>0$). Near the free-bound resonance, the excited state dominates and the position of free-ground resonance is to the blue, with a nodal position that depends on the detuning and intensity of the free-bound laser, as well as the the semi-classical size of both molecular states. The excited state contribution is to the blue because it takes extra energy to break up the excited-state molecule.

\begin{figure}[b]
\centering
\includegraphics[width=9.5cm]{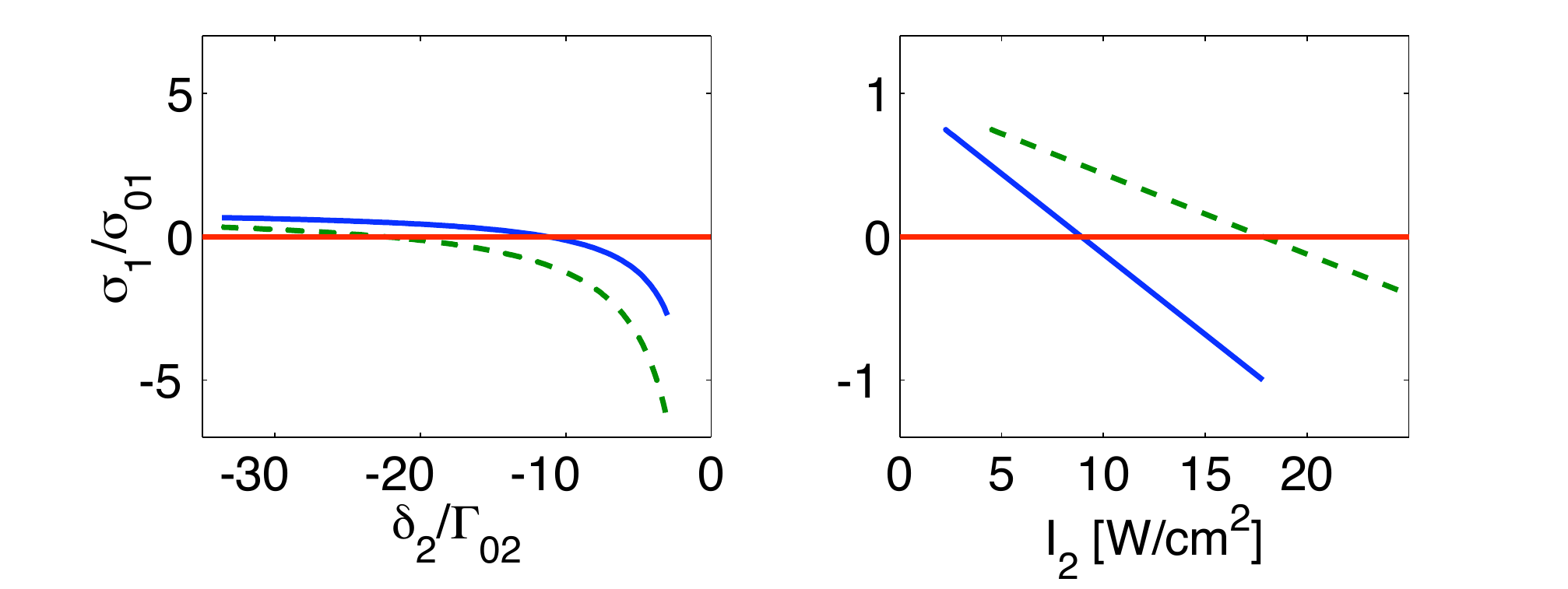}
\caption{Dispersive-like lightshift in heteronuclear free-ground photoassociation of a joint Li-Na Bose-Einstein condensate, where the free-ground laser intensity is 1~W/cm$^2$ and the size of the ground molecular state is estimated as $L_1/L_2=10$. (a) The detuning position of the lightshift node depends on the intensity of the free-bound laser, where the solid (dashed) line is for $I_2/I_1=100$~(200). (b) Similarly, the intensity position of the node depends on the free-bound detuning, where the solid (dashed) line is for $-\delta_2/\Gamma_{02}=1$~(2).}
\label{SHIFT}
\end{figure}

Quantitatively, up to an factor of $1/(\eta L_1/L)$, the light shift vanishes for a critical detuning given by Eq.~\eq{DET_NODE}, and a critical intensity given by Eq.~\eq{I_NODE}. Typical results for LiNa are shown in Fig.~\ref{SHIFT}. The difference in the position of the rate and lightshift nodes is due to the assumedly different sizes of the molecular states. In particular, the difference is due to the physical constraint that the size of the ground state is much less than the size of the excited state--if the roles are reversed, the positions of the nodes are roughly identical since $1/\eta\sim1$.

\subsection{Comparison to Collisional Models}

Bridging to collisional models, the laser-induced resonant scattering length is again defined by $\rho^{1/3}a_r=-\Omega_2^2/(8\pi\omega_\rho\delta_2)$. The rate constant~\eq{RES_RATE_CON} is then
\beq
\rho K=
  \frac{\Omega_1^2\left(1-\eta{\displaystyle \frac{a_r}{L}}\right)^2}
  {\Gamma_1+\left(\eta
   {\displaystyle\frac{a_r}{L}}
       {\displaystyle\frac{\Omega_1}{\Omega_2}}\right)^2
          \Gamma_2}\,.
\label{RES_RATE_CON_COLL}
\eeq
Similarly, the saturated rate constant, the saturation intensity, and the lightshift are re-written as
 \bml
\bea
\rho K&=&
  \left(1-\frac{1}{\eta}\,\frac{L}{a_r}\right)^2\frac{\Omega_2^2}{\Gamma_2}
\\
\frac{I_{01}'}{I_{01}}&=&
  \left(\frac{1}{\eta}\,\frac{L}{a_r}\frac{\Omega_2}{\Omega_{01}}\right)^2
     \frac{\Gamma_1}{\Gamma_2},
\\
\sigma_1&=&\sigma_{01}\left(1-\eta^2\frac{L_1}{L}\,\frac{a_r}{L}\right).
\eea
\eml

In this context, photoassociation ceases when the resonant scattering length matches the effective molecular size, $a_r\sim L$. As resonance is approached, $a_r\gg L$ and the system becomes strongly interacting--i.e., the free-ground rate constant saturates at the value of the free-bound rate constant--for smaller and smaller intensities $I_{01}'/I_{01}\ll1$. In the limit where the ground state is much smaller than the excited state, $L\sim L_2$, photoassociation ceases ($a_r\sim L_2$) and the net lightshift changes sign ($a_r\sim L_1/L_2^2$) at very different resonant scattering lengths since $L_1\ll L_2$. Lastly, the dependence on the free-bound laser intensity and detuning is now readily understood, since these are two means by which the scattering length can be tuned with photoassociation~\cite{FED96,MAC01,THE04}.

\subsection{Coherent Conversion}

Although a rate constant enhanced by three orders of magntiude is impressive, it is not necessarily the same as coherent conversion to the electo-vibrational ground state. Indeed, given that the enhanced free-ground rate constant saturates near-resonance at the rate for free-bound photoassociation alone, the question arises whether the conversion to ground state molecules is coherent, or whether the molecules decay as fast as--or faster than--they are made. Borrowing yet again from quantum optics, coherence is reached when the Rabi coupling is large compared to the damping rate, $|\chi|/\Gamma\gg1$. Using the collisional expression for its economy, the coherence factor is
\beq
\frac{\chi}{\Gamma}=
  \frac{\Omega_1\left(1-\eta{\displaystyle \frac{a_r}{L}}\right)}
  {\Gamma_1+\left(\eta
   {\displaystyle\frac{a_r}{L}}
       {\displaystyle\frac{\Omega_1}{\Omega_2}}\right)^2
          \Gamma_2}\,.
\label{CHI_O_GAMMA}
\eeq

\begin{figure}[t]
\centering
\includegraphics[width=4.25cm]{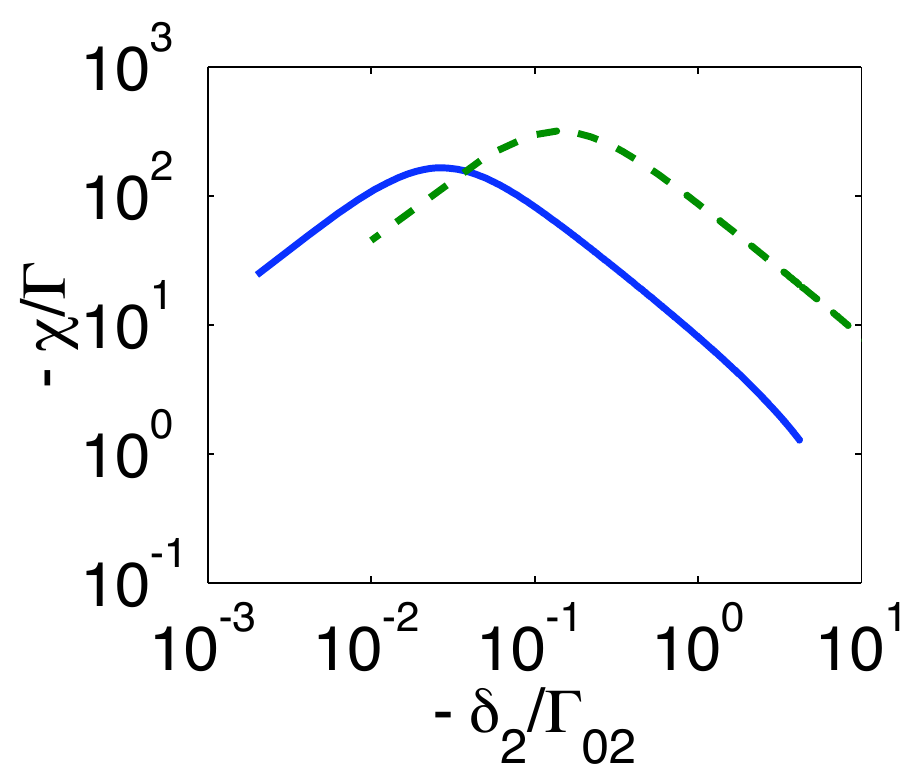}
\caption{Coherence in enhanced heteronuclear free-ground photoassociation is reached when the effective Rabi coupling is much larger than the effective decay rate, say, $|\chi|/\Gamma\sim10^2$. The free-ground intensity is fixed at 10~W/cm$^2$, and the solid (dashed) line is for $I_2/I_1=1$~(10).}
\label{CHI}
\end{figure}

Results for LiNa are shown in Fig.~\ref{CHI}. As resonance is approached, the coherence factor coupling increases and coherence is reached; but, it is eventually lost again, even as the rate constant continues to experience enhancement. According to Eq.~\eq{CHI_O_GAMMA}, the maximum occurs similar to saturation: for large free-bound detunings, $\Gamma_1$ dominates and $\chi/\Gamma\sim a_r$; for small detunings, $\Gamma_1$ is negligible and $\chi/\Gamma\sim1/a_r$. The inflection value of the scattering length is
\bml
\beq
\frac{a_r}{L}=\frac{1}{\eta}\,\frac{\Omega_2}{\Omega_1}\,
  \sqrt{\frac{\Gamma_1}{\Gamma_2}},
\eeq
which translates into an inflection detuning
\beq
\frac{\delta_c}{\Gamma_{02}}=-\frac{\eta}{8\pi}\,\frac{1}{\rho^{1/3}L}\,
  \frac{\Omega_1\Omega_2}{\omega_\rho\Gamma_{02}}\,\sqrt{\frac{\Gamma_2}{\Gamma_1}}.
\eeq
\eml
 The lowest detunings are on the verge of validity of the model but, again, we expect that these results are a reasonable approximation, nonetheless, based on intuition from the combined resonance system~\cite{MAC08}.

\section{Conclusion}
\label{CONCLUSION}

We have investigated ground-free-bound photoassociation of a heteronuclear Bose-Einstein condensate of atoms into dipolar molecules in the $J=1$ electro-vibrational ground state. Coupling the additional state to the continuum leads to a cross-coupling between the previously uncoupled molecules, which allows an alternate pathway for transitions to the ground molecular state and, therefore, quantum interference. While independent of the free-ground laser parameters, elimination and enhancement of the free-ground rate constant depends on the detuning and intensity of the free-bound laser, as well as the semi-classical size of both molecular states. Similarly, the cross-molecular coupling leads to a dispersive-like shift of the free-ground resonance position, which is also independent of the free-ground laser parameters, but does depend on the free-bound laser parameters and the size of both molecular states. Finally, although enhancement of the free-ground rate constant tops three orders of magnitude, there is only a small range in free-bound detuning where coherent conversion to ground state molecules is possible.

Besides an all-optical route to enhanced coherent conversion from heteronuclear atoms to dipolar molecules in the electro-vibrational ground state, the significance of the present work is elucidation of the role played by both molecular state sizes, and the subsequent contrast with the analogous Feshbach-enhanced photoassociation~\mbox{\cite{MAC08,PEL08,COU98,DEB09}}. In the case of Feshbach-enhanced free-bound transitions to an excited state~\cite{MAC08,COU98,DEB09}, the rate constant vanishes and the anomalous lightshift changes from red to blue at roughly the same magnetic field detuning, whereas here we find that rate constant vanishes and the anomalous lightshift changes sign at very different values. Based on the present results, this differential is due to an enhancing state that is much smaller (larger) than the target state in the magneto-optical (all-optical) system. Comparison to the magneto-optical heteronuclear case is difficult, since the results of Ref.~\cite{PEL08} are not explicit with respect to the size of the molecules. Nevertheless, the results herein provide a roadmap to a broader understanding of Feshbach-enhanced free-ground photoassociation to the heteronuclear electro-vibrational ground state.

\acknowledgements{
The authors gratefully acknowledge helpful conversations with Bill Stwalley, support from NSF (MM, 00900698), and a Diamond Research Scholarship from Temple University's Office of the Vice Provost for Undergraduate Affairs (CD).}


\begin{thebibliography}{99}
\bibitem{TEMP_NOTE}
The free-ground rate constant from Ref.~\cite{PEL08} has been scaled from the $T=50$~$\mu$K to the $T=100$~nK value according to $K_1\propto T^{-3/2}$. This assumes that $T\sim100$~nk is at or above the onset temperature of the Wigner threshold law, below which the rate constant is independent of temperature. If the onset temperature is higher, say, 500~nK, the free-ground intensities estimated herein will increase by about an order of magnitude.
\end{thebibliography}
\end{document}